\documentclass[runningheads]{llncs}
\usepackage{graphicx}
\usepackage{upgreek}
\usepackage{multirow}
\usepackage{url}
\usepackage[misc]{ifsym}

\begin{document}
\title{Lossy compression of multidimensional medical images using sinusoidal activation networks: an evaluation study}

\titlerunning{Lossy compression of multidimensional medical images}
%
\author{Matteo Mancini\inst{1} \and
Derek K. Jones\inst{1} \and
Marco Palombo\inst{1,2}\textsuperscript{(\Letter)}}
\authorrunning{M. Mancini et al.}
%
\institute{Cardiff University Brain Research Imaging Centre (CUBRIC), Cardiff University, Cardiff, UK - \email{\{mancinim,jonesd27,palombom\}@cardiff.ac.uk} \and School of Computer Science and Informatics, Cardiff University, Cardiff, UK}
\maketitle              
\vspace{-4mm}
\begin{abstract}
In this work, we evaluate how neural networks with periodic activation functions can be leveraged to reliably compress large multidimensional medical image datasets, with proof-of-concept application to 4D diffusion-weighted MRI (dMRI). In the medical imaging landscape, multidimensional MRI is a key area of research for developing biomarkers that are both sensitive and specific to the underlying tissue microstructure. However, the high-dimensional nature of these data poses a challenge in terms of both storage and sharing capabilities and associated costs, requiring appropriate algorithms able to represent the information in a low-dimensional space. Recent theoretical developments in deep learning have shown how periodic activation functions are a powerful tool for implicit neural representation of images and can be used for compression of 2D images. Here we extend this approach to 4D images and show how any given 4D dMRI dataset can be accurately represented through the parameters of a sinusoidal activation network, achieving a data compression rate about 10 times higher than the standard DEFLATE algorithm. Our results show that the proposed approach outperforms benchmark ReLU and Tanh activation perceptron architectures in terms of mean squared error, peak signal-to-noise ratio and structural similarity index. Subsequent analyses using the tensor and spherical harmonics representations demonstrate that the proposed lossy compression reproduces accurately the characteristics of the original data, leading to relative errors about 5 to 10 times lower than the benchmark JPEG2000 lossy compression and similar to standard pre-processing steps such as MP-PCA denosing, suggesting a loss of information within the currently accepted levels for clinical application. 

\end{abstract}

\section{Introduction}
Modern medical imaging provides both structural and functional information on anatomical features and physiological processes. In particular, Magnetic Resonance Imaging (MRI) is a formidable imaging technique providing a plethora of contrasts through different modalities which can be used to quantify specific features of biological tissues non-invasively. As such, the MRI use in clinics has transformed the diagnosis, management and treatment of disease.\\
Recent developments in both MRI scanners' hardware \cite{jones2018microstructural} and methods (e.g., \cite{palombo2020sandi,zhang2012noddi}) have pushed further the capabilities of medical imaging but also opened new challenges in terms of storage and sharing requirements of ever larger MRI datasets. For example, advanced 4D diffusion-weighted MRI (dMRI) datasets can require $\sim100\ MB$ - $10\ GB$, depending on the spatial resolution and the number of measurements. Moreover, large imaging studies scale this figure up to $>10\ TB$. An example is the UK-Biobank initiative (\url{https://www.ukbiobank.ac.uk}) which aims to collect extensive multi-modal MRI datasets from about 500,000 participants. For each participant, the size of a dMRI dataset is $550\ MB$, leading to a required storage of $275\ TB$ for the dMRI data alone.  Clearly, the high-dimensional nature of these data poses a challenge in terms of storage and sharing capabilities and associated costs and technology needs, requiring appropriate algorithms able to represent the information in a low-dimensional space.\\
In this respect, neural networks have been recently shown to be ideal tools to map pixel/voxel locations to image features. The learnt mappings are typically called implicit neural representations and have been used to represent images \cite{stanley2007compositional}, 3D scenes \cite{sitzmann2019scene}, videos \cite{li2021neural} audio \cite{sitzmann2020implicit}, and more. In particular, neural networks with periodic activation  functions have been recently proposed as powerful tools for implicit neural representation of images  \cite{sitzmann2020implicit} and can be used for efficient lossy compression of 2D images \cite{dupont2021coin}. With this strategy, there is no need to generalize to out-of-distribution samples: the compression procedure coincides with the training itself, with reasonable time/energy consumption on ordinary workstation or even low-power devices.
A drawback of this approach is that the resulting compression is lossy, meaning that the compression comes at the cost of losing a fraction of the information. The suitability of lossy compression in clinical applications has been widely investigated for CT, but relatively less for MRI. The few studies concerning MRI have concluded that JPEG and JPEG2000 lossy compression ratio up to 25 preserves diagnostic accuracy and perceived image quality \cite{terae2000wavelet}. However, some of the issues arising from the compression of CT data may be relevant to MR imaging. A wide range of studies have investigated mostly JPEG2000 compression of CT data and have concluded that acceptable lossy compression ratios range from 4 to 20 \cite{ohgiya2003acute,cosman1994thoracic,ko2003wavelet,yamamoto2001evaluation,lee2007irreversible}.\\
The aim of this work is to evaluate how neural networks with periodic activation  functions can be effectively leveraged to compress large multidimensional medical image datasets, with proof-of-concept application on 4D dMRI. By extending the approach proposed in \cite{sitzmann2020implicit,dupont2021coin} to 4D images, we quantitatively investigate the impact of lossy compression and show how any given 4D dMRI dataset can be accurately represented through the parameters of a sinusoidal activation network, achieving a data lossy compression rate about 10 times higher than the current standard lossless DEFLATE algorithm \cite{deutsch1996rfc1951}. We envision that this architecture will aid not only compression but also security and anonymization and could be applied to other image modalities, beyond this demonstration on dMRI.\\
\section{Methodology}
\begin{figure}[!t]
\centering
\includegraphics[width=0.85\linewidth]{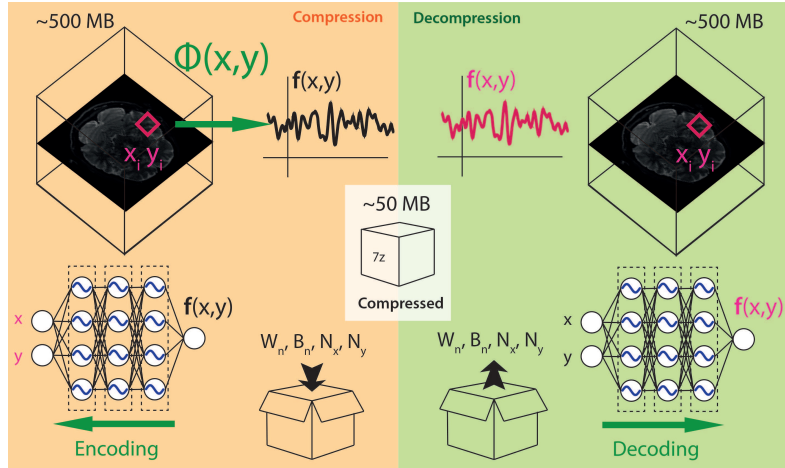}
\caption{\textbf{Compression and decompression procedure.} Overview of the main steps involved in the proposed compression and decompression procedure.} \label{fig1}
\end{figure}

The proposed compression approach uses the SIREN (sinusoidal representation networks) architecture \cite{sitzmann2020implicit}, which consists of a multilayer perceptron (MLP) with sine activation functions for implicit neural representations. The overall compression and decompression procedure is outlined in Fig.\ref{fig1}. As proposed in \cite{dupont2021coin}, the encoding step leverages the overfitting of the MLP to a given image, quantizing its weights and biases and storing those as a compressed reconstruction of the image. As a further lossless compression step, the network parameters are archived with a Lempel–Ziv–Markov chain algorithm (\url{https://7-zip.org}). At decoding time, the MLP is initialized with the stored parameters and evaluated using pixel locations as input to reconstruct the image.\\
We generalized this formulation for 4D datasets, where the goal becomes learning the mapping between the 3D coordinates of a given voxel and its modality-specific signal variations. As benchmark, we compared the performances of this approach with those from standard MLPs with ReLU and Tanh activation functions, widely used for implicit neural representation tasks \cite{sitzmann2020implicit}. To quantify their performances, we relied on several measures, including mean squared error (MSE), peak signal-to-noise ratio (PSNR) and structure similarity index (SSIM). Furthermore, we compared the loss of information from the proposed SIREN approach with benchmark lossy compression JPEG2000 in terms of relative error of estimated indices from tensor and spherical harmonics representations. Finally, as reference for the acceptable margins of these relative errors, we compared them with those resulting from two commonly used pre-processing steps in dMRI: MP-PCA denoising \cite{veraart2016denoising} and smoothing with a Gaussian kernel (FWHM=1.5).

\textbf{SIREN mathematical formulation.} We propose to approach the compression problem of a multi-dimensional image as finding an implicit representation \(\mathbf{\Upphi}\) between its coordinates \(\mathbf{x}_{i}\) and its features \(\mathbf{y}_{i} = \mathbf{f}(\mathbf{x}_{i})\). In the application to dMRI images mentioned in the introduction, each voxel is associated with a distribution of $N_{meas}$ directional values. To approximate \(\mathbf{\Upphi}\), we seek to solve the optimization problem with loss function \(\mathcal{L} = \sum{\| \mathbf{\Upphi}(\mathbf{x}_{i}) - \mathbf{f}(\mathbf{x}_{i})\|^2} \). Leveraging previous work on implicit representation, we define \(\mathbf{\Upphi}\) through a neural network with periodic activation function \cite{sitzmann2020implicit}:
\begin{equation}
\mathbf{\Upphi}(\mathbf{x}_{i}) = \mathbf{W}_{n}(\phi_{n-1} \circ \phi_{n-2} \circ \phi_{0})(\mathbf(x)) + \mathbf{b}_{n}, \qquad \mathbf{x}_{i} \mapsto \phi_{i}(\mathbf{x_{i}}) = sin(\mathbf{W}_{i}\mathbf{x}_{i}+\mathbf{b}_{i})
\end{equation}
where \(\phi_{i}\) represents the \(i^{th}\) layer of the network, \(\mathbf{W}_{i}\) is the weight matrix, and \(\mathbf{b}_{i}\) is the bias term. The main advantage of periodic activation functions is that the derivatives remain well behaved for any weight configuration, and as a result it is possible to learn not only the mapping \(\mathbf{\Upphi}\) but also all its derivatives. One caveat of these activation functions is that due to their periodic nature, catastrophic forgetting phenomena during training can occur. This issue can be easily overcome initializing the weights from a uniform distribution between \(-\sqrt(6/n)\) and \(\sqrt(6/n)\), where \(n\) is the number of inputs to each activation unit: this constrain ensures that the periodic activation input has a normal distribution with a unitary standard deviation \cite{sitzmann2020implicit}. In principle, this formulation can be applied to a tridimensional volume where \(\mathbf{x}_{i} = (x_{i}, y_{i}, z_{i})\). However, in this case the number of parameters needed to properly learn the mapping \(\mathbf{\Upphi}\) could become dramatically higher and poses a burden on the subsequent implementation. For this reason, a more feasible approach could be to learn \(\mathbf{\Upphi}\) for each slice relying on its bidimensional coordinates \(\mathbf{x}_{i} = (x_{i}, y_{i})\). Here, we implemented and compared both the approaches. A further observation is that combining SIREN and ReLU approches could lead to the best of both worlds, hence we also experimented with hybrid architectures.

\textbf{Implementation.} The networks were implemented using PyTorch, extending previous work made available from Sitzmann and colleagues (\url{https://github.com/vsitzmann/siren}). All the code is publicly available at the following GitHub repository: \url{https://github.com/palombom/SirenMRI}. For our experiments, we used five subjects from the publicly available MGH HCP Adult Diffusion dataset (\url{https://db.humanconnectome.org/}). The model fitting was performed using a NVIDIA Titan XP GPU for 2D architectures, while for 3D ones up to four NVIDIA Tesla V100 SXM2 GPUs were used in parallel. The training is performed in a self-supervised way, so the training set consists of all the voxels within a dMRI dataset (in this case 140x140x96=1881600). There is no split in training/validation/testing sets (and no need for it) as our goal is to overfit the input data. A new network is trained for each dataset. During training, which corresponds to the encoding or compression phase, we update the network’s weights and biases using back-propagation and mean squared error as loss function (calculated between the input data and the network’s predictions) which is minimized using the ADAM algorithm (2000 epochs, determined experimentally as trade-off between highest peak SNR, highest compression ratio and fastest training). The learning rates were determined experimentally as \(3\cdot10^{-4}\) and \(2\cdot10^{-4}\) for 2D and 3D architectures, respectively. The training (i.e. compression) for the whole dMRI dataset took on average 13 minutes, while the prediction (i.e. decompression) took about 2 seconds. We explored the impact of both deeper and wider architectures testing networks with 3/4/5 layers and 128/256/512 units per layer (1024 units for 3D architectures).

\textbf{Assessing networks' performances.} In the first experiment, we assessed the performance of each network. We quantified the average MSE, PSNR and compression ratio for a representative dMRI dataset (Subject ID: MGH1001) as:
\begin{equation}
MSE = \frac{1}{N_{voxels}} \frac{1}{N_{meas}} \sum_{i=1}^{N_{voxels}} \sum_{j=1}^{N_{meas}} [\hat{S}_j^{ground-truth}(\mathbf{x}_{i}) - \hat{S}_j^{decomp}(\mathbf{x}_{i})]^2
\end{equation}
\begin{equation}
PSNR = 20\ log_{10}(\frac{1}{\sqrt(MSE)}) 
\end{equation}
where $\mathbf{\hat{S}}(\mathbf{x}_{i})$ is the vector of $N_{meas}$ dMRI signals from the voxel at location $\mathbf{x}_{i}$, normalized between 0 and 1.
For each decompressed image, we also calculated voxel-wise the relative error and the local SSIM with respect to the ground-truth using the windowing approach proposed in \cite{wang2004image} and implemented in scikit-image (\url{https://scikit-image.org}). We finally computed the compression ratio as the ratio between the uncompressed and compressed image file sizes.

\textbf{Evaluating compression quality and accuracy.} In the second experiment, we quantitatively assessed the quality of the compression obtained by the network comprised of 3 layers and 256 units which showed a good compromise between PSNR ($>36$) and compression ratio ($\sim10$). Specifically, we quantified the accuracy of the compression for metrics of interest in dMRI applications, such as the diffusion tensor \cite{le2001diffusion}, the spherical harmonics coefficients \cite{descoteaux2007regularized} and the fibre orientation distribution function (fODF) \cite{tournier2007robust}. 
From each of the five subjects in our dataset, the diffusion tensor and related rotational invariant metrics fractional anisotropy (FA) and mean diffusivity (MD) were estimated using only the shell at $b=1,000\ s/mm^2$ with MRtrix3 (\url{https://www.mrtrix.org}) \cite{tournier2019mrtrix3}. The spherical harmonics coefficients up to the 4-th order were estimated using only the shell at $b=5,000\ s/mm^2$ with MRtrix3. The corresponding rotational invariant metrics RISH0 and RISH2 were then computed according to \cite{mirzaalian2016inter}. Finally, the fODF was estimated using the constrained spherical deconvolution algorithm as implemented in MRtrix3 using only the shell at $b=5,000\ s/mm^2$. 

\begin{figure}[!t]
\centering
\includegraphics[width=0.85\linewidth]{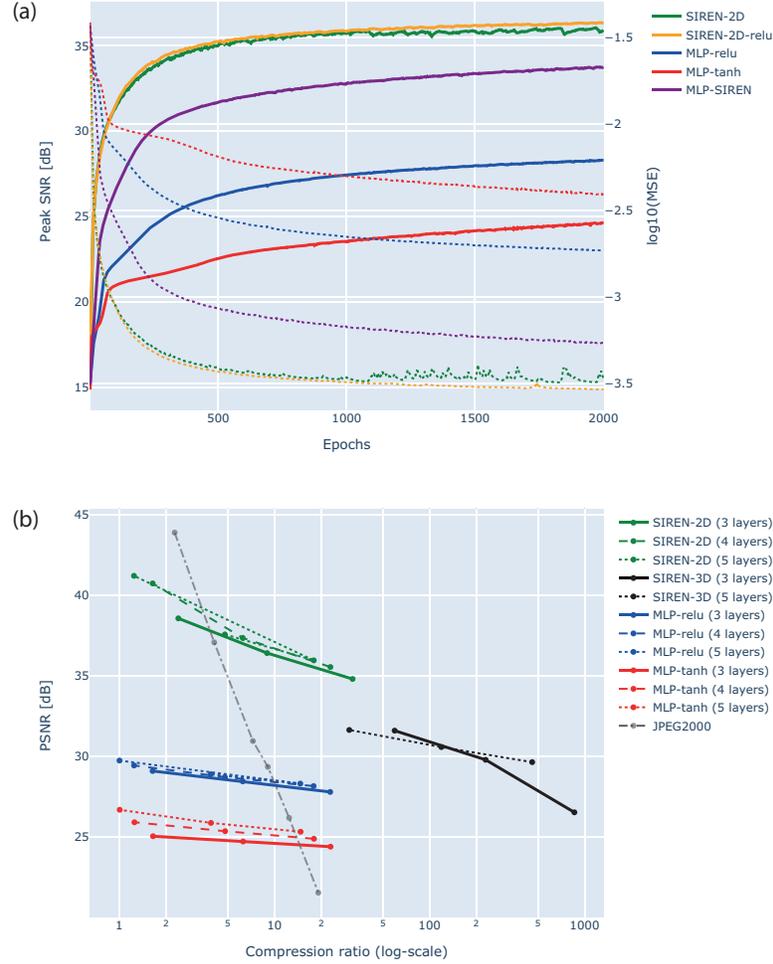}
\caption{\textbf{Networks' performances}. \textit{(a)} PSNR (solid lines) and log10(MSE) (dashed lines) as a function of training epochs for the different networks and the exemplar configuration of 3 layers and 256 units per layer, chosen as overall good compromise between the achievable PSNR and corresponding compression ratio. \textit{(b)} PSNR as a function of compression ratio for the different networks used and JPEG2000. In each curve, the points represent an increasing number of units per layer when going from higher to lower compression ratios. Specifically, for the 2D networks (SIREN-2D, MLP-ReLU and MLP-Tanh) these correspond to: 128/256/512 units, while for the 3D one (SIREN-3D): 256/512/1024 units.} \label{fig2}
\end{figure}

\section{Results}
\textbf{Networks' performances.} In Fig.\ref{fig2}a we show the performances during training for the representative networks comprised of 3 layers and 256 units, demonstrating that all the networks converged to their best PSNR/MSE. We note that introducing a ReLU activation function after the last layer of a SIREN-2D network (SIREN-2D-relu) stabilizes the training and slightly improves the performances, while adding a first layer with sinusoidal activation functions to a MLP with ReLU activation functions (MLP-SIREN) significantly improves the performances with respect to the ReLU-only MLP (MLP-relu) but still performing substantially worse than SIREN-2D and SIREN-2D-relu. In Fig.\ref{fig2}b we show the average slice-wise PSNR as a function of the compression ratio for each network. We observe that SIREN-2D outperforms the other 2D-based MLPs in terms of both PSNR and compression ratio; and the JPEG2000 at compression ratios higher than 3. With respect to the 3D implementation SIREN-3D, SIREN-2D still provides higher PSNR, although at a reduced compression ratio.

\setlength{\textfloatsep}{10pt}
\begin{figure}[!t]
\centering
\includegraphics[width=0.85\textwidth]{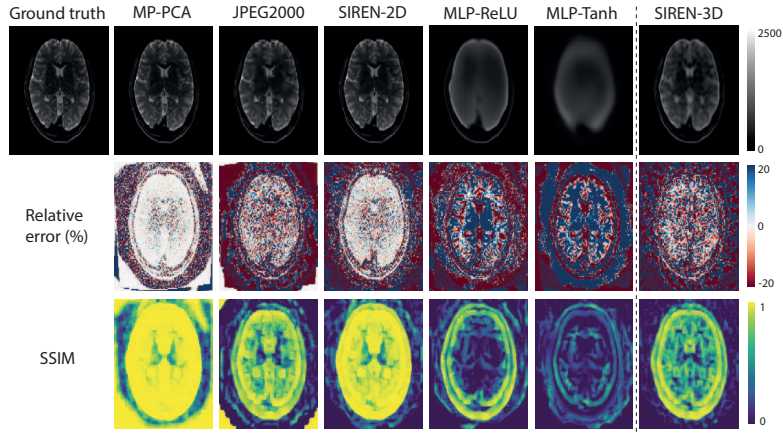}
\caption{\textbf{Compression quality of an exemplar dataset (MGH1001)}. The first row shows a representative slice of the b=0 image for the ground-truth and each decompressed image using JPEG2000 and the 2D (SIREN-2D, MLP-ReLU and MLP-Tanh) and 3D (SIREN-3D) networks, all with 3 layers and 256 units. As reference, the results after a commonly used pre-processing step - the MP-PCA denoising - are also reported. The second and third rows report the relative percentage error and the SSIM with respect to the ground-truth. The average PSNR and compression ratio are respectively: 29.3 and 9.0 for JPEG2000; 36.4 and 9.0 for SIREN-2D; 28.4 and 6.2 for MLP-ReLU; 24.7 and 6.3 for MLP-Tanh; 31.6 and 59.5 for SIREN-3D.} \label{fig3}
\end{figure}

\begin{table}[!t]
\caption{Relative error of diffusion-based metrics (MD: mean diffusivity and FA: fractional anisotropy from the data at b = 1000 $s/mm^2$; RISH: rotiationally invariant spherical harmonics from the data at b = 5000 $s/mm^2$) for one representative subject after the two exemplar pre-processing steps: MP-PCA denoising and smoothing with a Gaussian kernel; and compression/decompression using two lossy compression methods: the SIREN-2D network with 3 layers and 256 units, and the JPEG2000 algorithm with equivalent compression ratio. The mean ($\pm$ std) values were calculated using the voxels in the white (WM) and gray (GM) matter, and cerebrospinal fluid (CSF) masks.}\label{tab1}
\begin{tabular}{| c  c | c c c |}
\hline
 \multicolumn{2}{|c|}{\multirow{2}{*}{\textbf{Method}}} & \multicolumn{3}{|c|}{\textbf{MD}}\\
 & & \textit{WM} & \textit{GM} & \textit{CSF}\\
\hline
\multirow{2}{*}{\textit{Pre-Processing:}} & \textit{MP-PCA} & $1.98\% (\pm12.54\%)$ & $0.75\% (\pm1.27\%)$ & $1.64\% (\pm2.55\%)$\\

& \textit{Smoothing} & $3.02\% (\pm44.87\%)$ & $2.89\% (\pm13.03\%)$ & $-2.29\% (\pm3.73\%)$\\
\hline
\multirow{2}{*}{\textit{Compression:}} & \textit{JPEG2000} & $5.71\% (\pm132.38\%)$ & $3.89\% (\pm29.59\%)$ & $-4.01\% (\pm10.22\%)$\\

& \textit{SIREN-2D} & $2.16\% (\pm83.81\%)$ & $1.02\% (\pm7.45\%)$ & $1.56\% (\pm3.31\%)$ \\
\hline
\hline
\multicolumn{2}{|c|}{\multirow{2}{*}{\textbf{Method}}} & \multicolumn{3}{|c|}{\textbf{FA}}\\
& & \textit{WM} & \textit{GM} & \textit{CSF}\\
\hline
\multirow{2}{*}{\textit{Pre-Processing:}} & \textit{MP-PCA} & $0.44\% (\pm4.58\%)$ & $-0.12\% (\pm9.10\%)$ & $-1.10\% (\pm18.05\%)$\\

& \textit{Smoothing} & $-5.36\% (\pm5.75\%)$ & $-8.68\% (\pm8.34\%)$ & $-8.49\% (\pm11.18\%)$\\
\hline
\multirow{2}{*}{\textit{Compression:}} & \textit{JPEG2000} & $-7.98\% (\pm21.92\%)$ & $-0.09\% (\pm36.63\%)$ & $5.39\% (\pm40.73\%)$\\

& \textit{SIREN-2D} & $-0.72\% (\pm9.82\%)$ & $-1.03\% (\pm16.13\%)$ & $-0.07\% (\pm22.81\%)$\\
\hline
\hline
\multicolumn{2}{|c|}{\multirow{2}{*}{\textbf{Method}}} & \multicolumn{3}{|c|}{\textbf{RISH0}}\\
& & \textit{WM} & \textit{GM} & \textit{CSF}\\
\hline
\multirow{2}{*}{\textit{Pre-Processing:}} & \textit{MP-PCA} & $0.01\% (\pm0.80\%)$ & $0.01\% (\pm0.72\%)$ & $0.01\% (\pm0.66\%)$\\

& \textit{Smoothing} & $-1.83\% (\pm2.63\%)$ & $0.78\% (\pm5.24\%)$ & $6.16\% (\pm7.99\%)$\\
\hline
\multirow{2}{*}{\textit{Compression:}} & \textit{JPEG2000} & $-1.54\% (\pm5.24\%)$ & $0.69\% (\pm11.89\%)$ & $11.02\% (\pm21.09\%)$\\

& \textit{SIREN-2D} & $-0.50\% (\pm3.39\%)$ & $0.24\% (\pm5.31\%)$ & $0.99\% (\pm7.44\%)$\\
\hline
\hline
\multicolumn{2}{|c|}{\multirow{2}{*}{\textbf{Method}}} & \multicolumn{3}{|c|}{\textbf{RISH2}}\\
& & \textit{WM} & \textit{GM} & \textit{CSF}\\
\hline
\multirow{2}{*}{\textit{Pre-Processing:}} & \textit{MP-PCA} & $-0.73\% (\pm10.91\%)$ & $-6.94\% (\pm41.90\%)$ & $-10.95\% (\pm45.00\%)$\\

& \textit{Smoothing} & $-11.41\% (\pm8.18\%)$ & $-19.78\% (\pm16.17\%)$ & $-27.42\% (\pm21.47\%)$\\
\hline
\multirow{2}{*}{\textit{Compression:}} & \textit{JPEG2000} & $-8.60\% (\pm18.78\%)$ & $-6.32\% (\pm62.54\%)$ & $10.25\% (\pm106.32\%)$\\

& \textit{SIREN-2D} & $-4.92\% (\pm17.55\%)$ & $-6.80\% (\pm53.25\%)$ & $2.52\% (\pm92.75\%)$\\
\hline
\end{tabular}
\vspace{5mm}
\end{table}

\begin{table}[!t]
\caption{Relative error for five subject after compression/decompression of diffusion-based metrics (MD and FA from the data at b = 1000 $s/mm^2$; RISH from the data at b = 5000 $s/mm^2$) for the SIREN-2D network with 3 layers and 256 units. The mean ($\pm$ std) values were calculated using all the voxels in the brain mask, including CSF.}\label{tab2}
\begin{tabular}{|c | c c | c c|}
\hline
 \textbf{Subject ID} & \textbf{MD} & \textbf{FA} & \textbf{RISH0} & \textbf{RISH2}\\
\hline\hline
\textit{MGH1001} & $-0.69\% (\pm5.37\%)$ & $-3.93\% (\pm11.96\%)$ & $-0.06\% (\pm4.54\%)$ & $-13.99\% (\pm28.00\%)$\\
\hline
\textit{MGH1002} & $-0.72\% (\pm5.57\%)$ & $-3.89\% (\pm12.17\%)$ & $-0.11\% (\pm4.66\%)$ & $-14.17\% (\pm28.14\%)$\\
\hline
\textit{MGH1003} & $-0.87\% (\pm4.99\%)$ & $-4.37\% (\pm12.06\%)$ & $-0.13\% (\pm4.72\%)$ & $-15.56\% (\pm29.64\%)$\\
\hline
\textit{MGH1004} & $-0.82\% (\pm5.48\%)$ & $-3.94\% (\pm12.32\%)$ & $-0.09\% (\pm4.60\%)$ & $-14.51\% (\pm29.04\%)$\\
\hline
\textit{MGH1005} & $-0.65\% (\pm5.23\%)$ & $-3.65\% (\pm11.56\%)$ & $-0.10\% (\pm4.49\%)$ & $-11.33\% (\pm25.60\%)$\\
\hline
\end{tabular}
\end{table}

\begin{figure}[!t]
\centering
\includegraphics[width=0.9\linewidth]{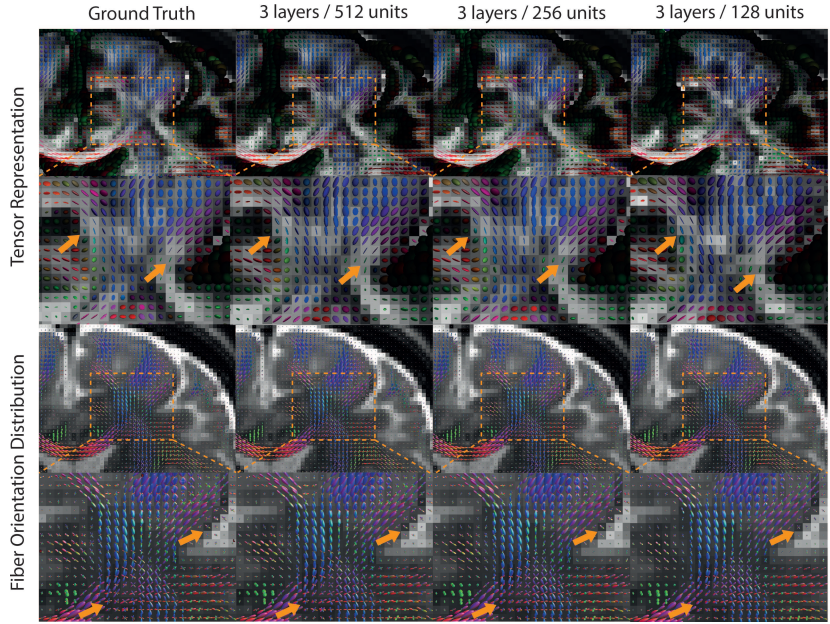}
\caption{\textbf{Reconstructed diffusion
tensor and fODF for an exemplar dataset (MGH1001) and different SIREN-2D architectures}. Arrows point at regions where differences are more evident, especially for the 3 layers/128 units case.}
\label{fig4}
\end{figure}

\textbf{Compression quality and accuracy.} In Fig.\ref{fig3} we compare the compression quality for the commonly used lossy compression JPEG2000 and all the networks comprised of 3 layers and 256 units using the b=0 image from the subject MGH1001. We find that SIREN-2D provides the lowest relative error and the highest SSIM among the lossy compressions. It is worthwhile noting that masking the brain, if the background is not of interest, can lead to further compression of a factor 2 to 3. In terms of diffusion metrics, in Tab.\ref{tab1} we report the comparison of the relative error obtained by SIREN-2D with 3 layers and 256 units with the relative error obtained by JPEG2000 and two pre-processing steps commonly used in dMRI data analysis: MP-PCA denoising and Gaussian smoothing.We found that the relative error from SIREN-2D is close to the relative error from MP-PCA denoising, and from about 5 to 20 times lower than smoothing and JPEG2000 compression. In Tab.\ref{tab2} we report the relative error after compression and subsequent decompression for FA, MD, RISH0 and RISH2, obtained with the same SIREN-2D network with 3 layers and 256 units across five subjects. Consistently, we find that the relative error on FA is $<5\%$, on MD and RISH0 $<1\%$ and on RISH2 $\sim15\%$.
Finally, in Fig.\ref{fig4} we show the impact of the number of units for the SIREN-2D network with 3 layers on the diffusion tensor and the fODF. As expected, SIREN-2D with 512 units provides the most accurate result, however its compression ratio is only $\sim2$. Very small differences can be seen between this network and SIREN-2D with 256 units, while some relevant differences are noticeable with SIREN-2D with 128 units.

\setlength{\textfloatsep}{10pt}

\setlength{\parskip}{2pt}
\textbf{Summary.} Overall, our results show that SIREN-2D with 3 layers/256 units is a good compromise to achieve a compression ratio of $\sim10$ with high accuracy. Importantly, the relative error obtained with this approach is a) smaller than commonly used lossy JPEG2000 compression; b) much smaller than commonly used pre-processing steps of Gaussian smoothing and c) similar to the commonly used pre-processing MP-PCA denoising, suggesting minimal loss of information within the currently accepted levels for clinical applications and utility. 

\section{Discussion and Conclusion}

The results of our experiments show how periodical activation functions can dramatically change the performances of MLPs, providing a powerful tool for compression purposes. This is thanks to the ability of periodical activation functions to learn the non-linear mapping between the position of each voxel in a 4D dMRI image and the corresponding features of the dMRI signal \textit{as well as} all its derivatives \cite{sitzmann2020implicit}. Here, we have demonstrated two important aspects for the design and training of sinusoidal activation networks: a) a single layer with sinusoidal activation functions in a conventional MLP with ReLU (or Tanh) activation functions improves its performances but is not enough to fully exploit SIREN's properties; b) a ReLU activation function after the last layer of a sinusoidal activation network provides more stable convergence to the optimum.

This work is inspired by recent studies showing how SIREN networks can be used to compress several kind of data, including images from different modalities, videos and audios \cite{dupont2022coinpp,sitzmann2020implicit,dupont2021coin}. However, none of the previous studies have assessed quantitatively the downstream impact of the SIREN lossy compression on medical images analysis, and to what extent this is acceptable. To the best of our knowledge, this is the first work to quantitatively evaluate the performance of SIREN for compression of multidimensional medical imaging modalities. Our quantitative results highlight how the compression procedure, despite lossy, preserves the measures underlying the images with high accuracy. In fact, the estimated relative error subsequent to SIREN compression (up to compression ratio 10) were very similar to the error from commonly used pre-processing steps, such as MP-PCA denoising. This suggests minimal loss of information within the currently accepted levels for clinical applications and utility.

It is worthwhile noting that, for compression ratios $\leq2$, JPEG2000 outperforms SIREN-2D(-relu) in terms of peak SNR and it would be a better choice for the lossy compression. For higher compression ratios, SIREN-2D(-relu) should be used instead, as it largely outperforms JPEG2000. Also, on average, the lossy compression algorithms tested here showed large standard deviations which may require further investigations. With respect to JPEG2000, SIREN-2D showed consistently lower standard deviations, but larger than MP-PCA denoising. 

In addition to the good compromise between PSNR and compression ratio, the proposed implementation does not require to generalize as the goal of this approach is to overfit the data itself. Therefore it does not require large datasets for training and testing, as in other deep learning based compression methods (e.g. autoencoders \cite{theis2017lossy}). Our results from different subjects confirm the stability of the compression performances. Moreover, given the relatively shallow and simple architecture of the SIREN network, the proposed method can be implemented on ordinary workstations, offering a valid low-energy/low-cost alternative to more demanding deep learning solutions. Future iterations could take advantage of multiple low-energy devices to parallelize the compression, reducing even more the associated costs. In fact, the main cost behind this approach is related to the time consumption of the compression phase, e.g. 8 seconds per slice with the 3-layers/256-units SIREN-2D on a NVIDIA Titan XP GPU (our datasets are comprised of 96 slices). Nonetheless, in the application scenario we envision, the compression needs to be done only once and could potentially be done overnight, even directly on the MRI scanner workstation.

It is finally worth noting that, as most multi-dimensional medical imaging modalities are inherently tridimensional, a 3D architecture would be a tempting choice. However, as observed in our experiments, the number of parameters necessary to obtain reasonable PSNR would be prohibitive: even using multiple GPUs in parallel, it was not possible to outperform the simpler 2D implementation. We refer the reader to recent works explaining further the challenges involving 3D compression and possible solutions \cite{dupont2022coinpp,mehta2021modulated}. Potential future directions could explore patch-based approaches, as indeed recently explored in these works. Future work will also explore the performance of the proposed compression methods on multidimensional MRI data entailing the acquisition of diffusion images with different TE, TR, etc, such as the MUDI dataset \cite{pizzolato2020acquiring}.

\setlength{\parskip}{5pt}
\noindent\textbf{Acknowledgements.} MM is supported by the Wellcome Trust through a Sir Henry Wellcome Postdoctoral Fellowship (213722/Z/18/Z). MP is supported by the UKRI Future Leaders Fellowship MR/T020296/2.

%
%
%
%
\bibliographystyle{splncs04}
\bibliography{refs}
\end{document}